\documentclass[twocolumn,aps,pra,superscriptaddress,showpacs]{revtex4-1}
\usepackage{graphicx}
\usepackage[version=3]{mhchem} 
\usepackage[usenames,dvipsnames]{xcolor}
\begin{document}

\title{Fully saturated hydrocarbons as hosts of optical cycling centers}
\author{Claire E. Dickerson}
\author{Cecilia Chang}
\author{Han Guo}
\author{Anastassia N. Alexandrova}
\email{ana@chem.ucla.edu}
\affiliation{Department of Chemistry and Biochemistry, University of California, Los Angeles, California 90095, USA}
\date{\today}

\begin{abstract}
Designing closed, laser-induced optical cycling transitions in trapped atoms or molecules is useful for quantum information processing, precision measurement, and quantum sensing.  Larger molecules that feature such closed transitions are particularly desirable, as they extend the scope of applicability of such systems.  The search for molecules with robust optically cycling centers has been a challenge, and requires design principles beyond trial-and-error.  Here, two design principles are proposed for the particular architecture of M-O-R, where M is an alkaline earth metal radical, and R is a ligand:  1) Fairly large saturated hydrocarbons can serve as ligands, R, due to a substantial HOMO-LUMO gap that encloses the cycling transition, so long as the R group is rigid.  2) Electron-withdrawing groups, via induction, can enhance Franck-Condon factors (FCFs) of the optical cycling transition, as long as they do not disturb the locally linear structure in the M-O-R motif.   With these tools in mind, larger molecules can be trapped and used as optical cycling centers, sometimes with higher FCFs than smaller molecules.
\end{abstract}

\maketitle

\section{Introduction}
Molecules which can be laser cooled and favor electronic transitions between only a few controlled, electronic states (optical cycling centers, or OCCs) are useful for the fields of quantum information processing, magnetic sensing, and ultracold physics studies.\cite{intro1,intro2,intro3,intro4,intro5,intro6,intro7,intro8,intr09}  Using trapped-ion hardware and theory from atomic physics, small molecules have been laser cooled and shown to exhibit these properties.\cite{Shuman2009DiatomicRadiativeForce,Hummon20132D, Truppe2017Molecules,Kozyryev2017Sisyphus,Augenbraun2020LaserCooled}  The success of these molecules relies on high Franck-Condon factors (FCFs), which in many cases is quite close to the vibrational branching ratios in these species.\cite{isaev2016polyatomic}

In the past, trapping and laser cooling molecules larger than a few atoms seemed formidable due to an increase in the number of vibrational modes, therefore a likely increase in vibrational branching.  This vibrational branching poses a threat to laser cooling of molecules for quantum information,\cite{DeMille2002Quantum,Yellin2006Schemes,Herrera2014InfraredDressed,Mallikarjun2016Prospects,Blackmore2019Ultracold,Ni2018Dipolar,Hudson2018Dipolar,Yu2019Scalable,Campbell2020DipolePhonon} as it would require many more repump lasers than experiment can afford.  Successful molecular candidates are those where the ground and excited state potential energy surfaces are highly parallel and their electronic transitions are within the Franck-Condon region.  Early investigations into this bonding structure began with alkaline earth molecules bonding to hydroxide and rigid, electronegative ligands such as borohydride and cyclopentadiene.\cite{Brazier1986,bernathcabh41990,ortiz1991electron,ortiz1991c5h5}

Previous work on optical cycling and laser cooling of molecules began with diatomics,\cite{shuman2010laser,anderegg2017radio} then M-O-H species (where M is the alkaline earth radical bonded to an oxygen),\cite{Brazier1986,brazvilas2022magneto} and recently, alkaline earth alkoxides, built upon an M-O-R scheme where R is a saturated hydrocarbon.\cite{Kozyryev2016Proposal,ivanov2019towards,Klos2020Prospects,Ivanov2020Toward, Augenbraun2020Molecular} However, which R ligands are suitable for these systems, i.e. what limitations there are on their size and chemical nature, remains an open research question.  Previous successful candidates were found through guesswork, rather than guided design.\cite{Klos2020Prospects,Ivanov2020Toward}  Most recently, multivalent OCCs were explored which required different design principles than the M-O-R framework.\cite{Hutzler}

Our past work focused on introducing design principles for M-O-R OCCs.  We proposed that it is possible to expand these saturated hydrocarbon ligands to much larger sizes, including diamond and cubic boron nitride.  The M-O-R motif with such large R is still a good optical cycling candidate, due to the ligand's large band gap.\cite{GuoSurfaceChemicalTrapping2020}  Additionally, we found that arenes could be suitable ligands for optical cycling groups (or quantum functional groups) so long as the arene's HOMO-LUMO gap is large enough to encapsulate the optical cycling transition.\cite{dickerson2021optical,Debayan2022}  In this work, we aim to strengthen these design principles by investigating other possible organic ligands of varying size and degree of electron-withdrawing character, which could work as optical cycling centers (OCCs), and emphasize how symmetry and ligand rigidity affects OCCs.  

Additionally, we will probe the effect of substituents on these R ligands, motivated by our recent study on alkaline earth phenoxides serving as good OCCs.\cite{dickerson2021franck,Guozhu2022CaOPh}  In particular, we showed we can strategically diagonalize FCFs by adding electron-withdrawing substituents onto the phenyl ring, enforcing closer geometries between ground to excited states and favoring fewer, more dominant vibrational modes contributing to off-diagonal decays.  This was due to the M-O bond becoming progressively more ionic, because of the increased pull in electron density.  While in aromatic systems the substituents operate through resonance, in this work, we show the effect still holds via induction, though to a weaker degree, in saturated R group ligands. 

Design rules in this paper can be applied to a wide variety of open-shell molecules, in which an increase in ligand complexity does not lower FCFs, but rather outcompetes simpler molecules through built-in electronic effects.  

\section{Theoretical Methods}
Molecule geometries were optimized with density functional theory (DFT) and time-dependent density functional theory for excited states (TD-DFT) at the PBE0-D3/def2-TZVPPD level of theory in Gaussian16.\cite{Perdew1996Rationale,D3-Grimme2010,Rappoport2010def2-tzvppd,g16}  An effective core potential (ECP) was used for the Sr atom.  Franck-Condon factors were then calculated using the harmonic approximation including Duschinsky rotations in Gaussian16.  Molecular orbitals were generated using Multiwfn.\cite{Multiwfn}  

Previous theoretical studies on OCCs used Complete Active Space Self-Consistent Field (CASSCF) and Multireference configuration interaction (MRCI) methods to produce highly accurate results. \cite{Hao2019High,Kang2016Suitability,Kozlov1997Enhancement,Nayak2006Ab,Tohme2015Theoretical}   However, scaling is poor with system size.  Hence, cheaper methods must be applied for large molecular screening.  Here, we find DFT/TD-DFT methods are a suitable alternative, within a margin of error, when picking the functional and basis set carefully.  

In previous work, we compared CaOH, SrOH geometries and vertical excitation energies calculated with CAS/MRCI to those computed with  TD-DFT, and both were compared to available experimental data.\cite{dickerson2021franck,Guozhu2022CaOPh}  At the PBE0-D3/def2-TZVPPD level of theory, TD-DFT-based bond lengths deviated from experimental values by less than 0.02 \AA  for both ground and excited states, while the computed CASSCF bond lengths are 0.06 \AA  longer than the TD-DFT values.  For larger molecules, such as CaOPh and CaO1Nap, our level of theory was benchmarked to experiment and found vertical excitation energies to be within 0.08 eV of experiment (see Table S1 in Supporting Information for details).\cite{Guozhu2022CaOPh,Debayan2022}  
Overall, we find DFT a suitable alternative to more expensive methods for predicting FCFs for these species, particularly when chemical trends are of interest more than the absolute accuracy, and will use it throughout this work. 

\section{Results and discussion}

\subsection{Hydrocarbon Ligands for OCCs}
First, we explore the saturated hydrocarbon manifold, for increasingly larger R group ligands.  The R group can extend to larger size so long as the ligand structure is rigid and spatially separated from the metal.  A related requirement is the local linearity of the M-O- motif, to avoid the bonding of the metal radical to parts of the molecular host.  Also, the ligand must have a larger HOMO-LUMO gap than the metal to metal OCC transition, and fully enclose the OCC transition, i.e. have the intrinsic HOMO of R below the M-based HOMO of M-O-R, and the intrinsic LUMO of R above the M-based LUMO of M-O-R.\cite{dickerson2021optical}

\begin{figure}
    \centering
    \includegraphics[width = 0.45\textwidth]{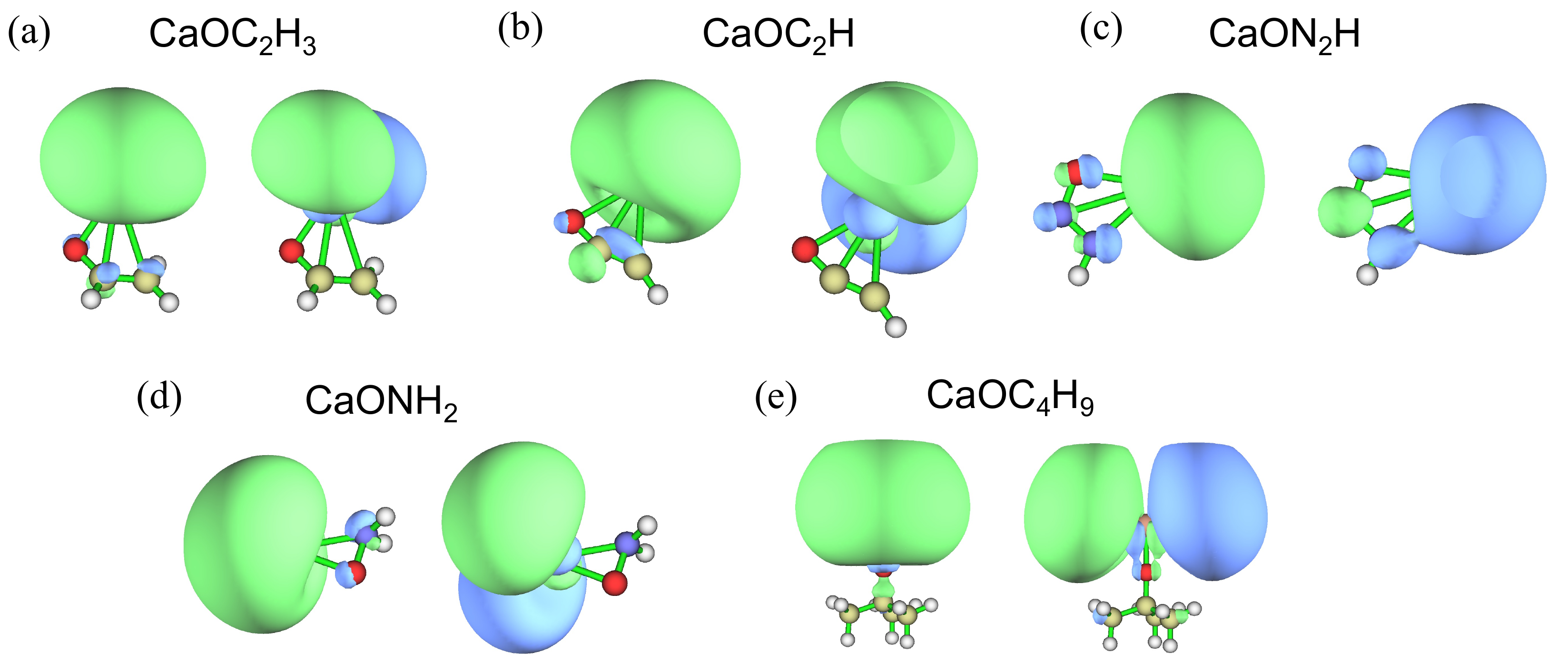}
    \caption{Highest occupied molecular orbitals (HOMOs) and lowest unoccupied molecular orbitals (LUMOs) for poor OCC candidates.}
    \label{fig:MO-fail}
\end{figure}

We investigated these limits in more detail.  We found that unsaturated non-aromatic hydrocarbons are poor ligands and will fail, because their $\pi$-system is much more reactive and will interact with the metal or oxygen, disrupting the clean metal-metal transition.  For example, as seen in Figure \ref{fig:MO-fail}, the HOMO of the ethenyl- or ethynyl- containing derivatives, the metal s and p orbitals mix with the O p orbital as well as the C-C $\pi$ orbital, indicating that the molecules are less ionic.  The LUMO is dominated by the metal p and d orbitals.  Thus, we expect the electronic transitions to couple to the molecular vibrations, giving rise to less diagonal FCFs.  Similar bonding and orbital mixing have been observed in M-O-R with the -N$_{2}$H ligand.  In the most stable geometry of the ground electronic state, the metal will bind to the O atom and the N-N $\pi$ bond near the metal, featuring charge density delocalization with the ligand.

Additionally, we find that even if a structure is locally linear, such as  CaOC$_{4}$H$_{9}$ which was previously confirmed experimentally,\cite{Brazier1986} it could still be a poor candidate for optical cycling.  In this case, the R ligand is too flexible, and the states on Ca have an overlap with the R ligand, such that the excited states have electron density on the CH$_{3}$ groups.  This coupling produces a quartic well in which there are two local minima on the excited state (see Figure S1 in the SI for additional details).  As a result, the diagonality of the ground to excited state transition is lost. 

\begin{figure}
\centering
  \includegraphics[width=0.45\textwidth]{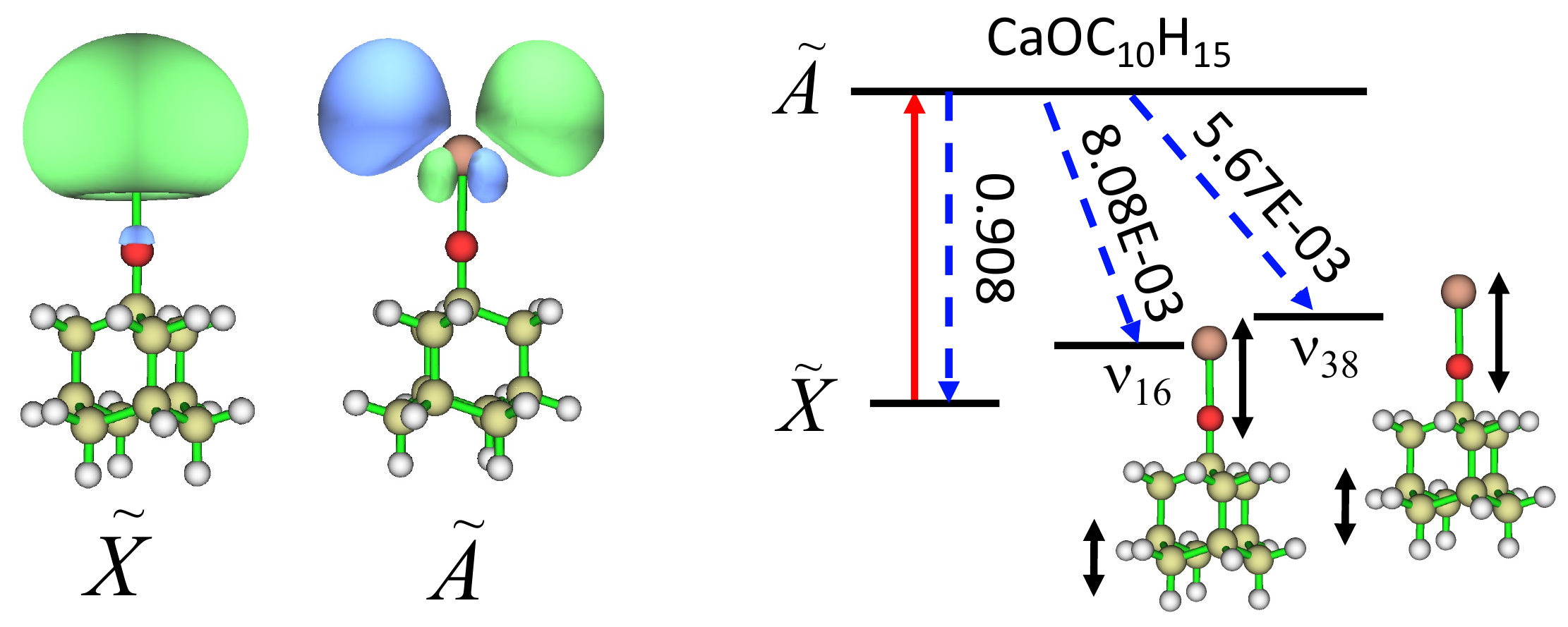}
  \caption{Natural transition orbitals (NTOs) for CaOC$_{10}$H$_{15}$'s ground ($\Tilde{X}$) to 1st ($\Tilde{A}$) excited state transitions.  Orbitals generated with an isosurface value of 0.03.  Additionally, excitation (red) and most dominant off-diagonal vibrational decays (blue) are shown, along with the associated Franck-Condon factors.}
  \label{fig:MO-other}
\end{figure}

Contrastingly, a molecule of the same $C_{3v}$ symmetry as CaOC$_{4}$H$_{9}$, CaO-adamantane, appears to be a good candidate for optical cycling.  This is due to the ligand rigidity which prevents ligand-Ca interaction.  Our calculations show that the electronic HOMO $\rightarrow$ LUMO transition in CaO-adamantane is quite isolated on the metal, as seen by the Natural Transition Orbitals (NTOs) in Figure \ref{fig:MO-other}.  The natural population analysis (NPA) of the charge populations are 0.838 for Ca and -1.069 for O for the ground state.  The transition orbitals resemble unhybridized $s$ and $p$ orbitals which are seen in the gas-phase atomic Ca$^+$ ion, further supporting the promise to use this molecule in optical cycling.

 The largest geometry change is the Ca-O bond length.  Ca-O-C is linear for these molecules in the $\Tilde{X}$ and $\Tilde{A}$ states.  The Ca-O bond length change from the $\Tilde{X}$ state to the $\Tilde{A}$ state is 0.023 $\AA$ in CaO-adamantane.

As seen by Figure \ref{fig:MO-other}, the most dominant off-diagonal decay for CaO-adamantane are dominated by stretching motions.  We speculate CaO-adamantane is a more successful candidate than CaOC$_{4}$H$_{9}$ because, despite the same symmetry, the large mass and rigidity of adamantane encourages the vibrational stretching motion, instead of the more flexible $C_{4}H_{9}$, which encourages CH$_{3}$ bending motions, which mix the metal Ca with the CH$_{3}$ groups, disrupting the isolated electronic transition.

Although the FCF for CaO-adamantane (0.908) is worse than a those of M-O-Rs with certain smaller ligands, such as phenyl (0.933\cite{Guozhu2022CaOPh}), the transition is not destroyed by adding a larger ligand.  In fact, it has a better FCF than smaller molecules, such as CaOC$_{2}$H$_{5}$ ($<0.90$\cite{paul2016dispersed}) and CaOCH(CH$_{3}$)$_{2}$ (0.720\cite{telfah2022combined}), despite an increase in the number of vibrational modes.  As such, this demonstrates that an increase in ligand complexity in some cases outcompetes simpler molecules, paving the way for larger optical cycling centers to be realized in the future.

In summary, saturated hydrocarbon molecules with large HOMO-LUMO gaps can be appended as ligands to alkaline earth oxides and still allow the M-O motif function as an optical cycling center.  As long as the ligand remains rigid, a large increase in a molecule's complexity can still produce a similar or better Franck-Condon factor than a smaller molecule.  

Additionally, symmetry is an important consideration for optical cycling center design.  Unsubstituted CaO-adamantane has $C_{3v}$ symmetry.  However, as we will now show, substituting hydrogens can reduce the symmetry and in some cases boost FCFs, if the substituted groups are electron-withdrawing.

\subsection{Inductive Electron-Withdrawing Groups Increase FCF}
\begin{figure}[ht]
\centering
  \includegraphics[width=0.45\textwidth]{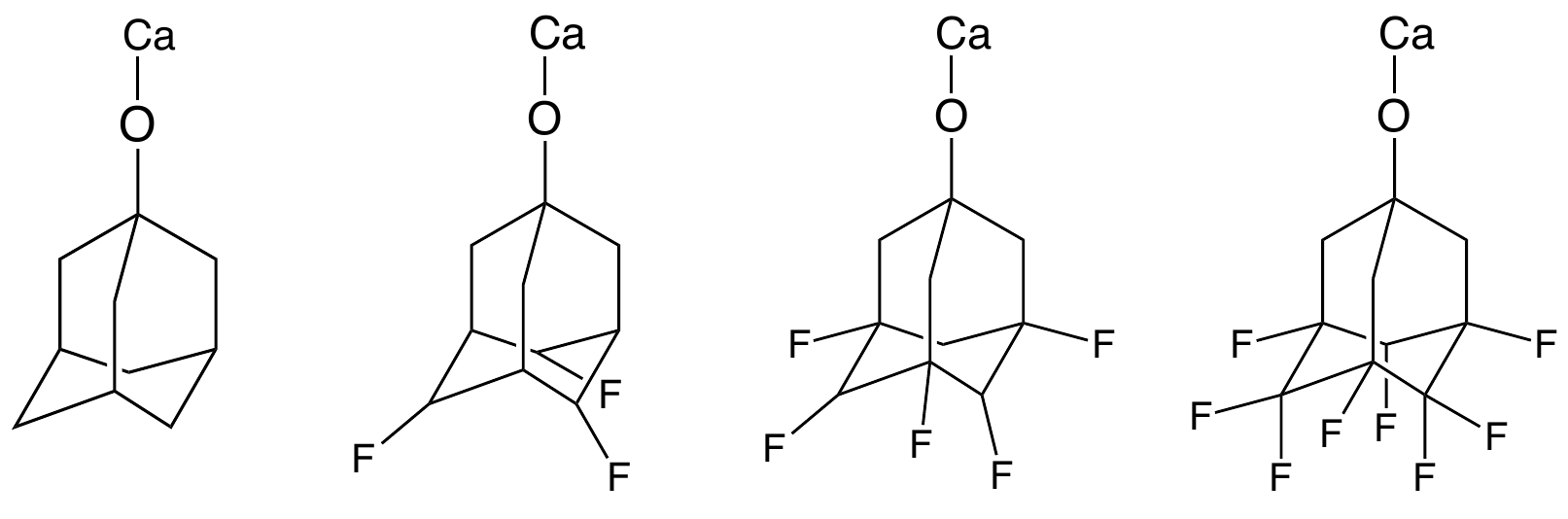}
  \caption{Molecules decorated with optical cycling motifs that are considered in this substitution study.  From left to right: CaO-Adamantane, CaO-3F-Adamantane, CaO-5F-Adamantane and CaO-8F-Adamantane.}
  \label{fig:substitutions}
\end{figure}

FCFs can also be altered by substituting electron-donating or electron-withdrawing groups.  Previously, we showed electron-withdrawing groups improved FCFs through resonance interactions, but the strength of this effect or the design rule for fully saturated hydrocarbons remains unexplored.\cite{dickerson2021franck,Guozhu2022CaOPh}  Here, we find that substituting electron-withdrawing groups increases FCFs inductively, through $\sigma$-bonding effects and makes the Ca-O bond more ionic.  Contrastingly, substituting electron-donating groups decreases FCFs.  We demonstrate this by substituting CaO-adamantane, replacing three, five, and eight hydrogen atoms at a time, in different positions, as shown in Figure   \ref{fig:substitutions}.  We also expand the scope toward SrO- variants (previously shown to have systematically less diagonal FCFs compared to CaO-, in the form of phenoxides).\cite{dickerson2021franck} 

Adding electron-withdrawing groups to both CaO- and SrO-adamantane  increases FCFs through $\sigma$-bonding effects, making the Ca(Sr)-O bond more ionic, as can be seen from NPA charges on the Ca(Sr) and O atoms (see Tables \ref{tab:ca-ad} and \ref{tab:subad-FCF}, ordered according to the electron-withdrawing/donating strength of the substituents).  For example, the unsubstituted CaO-adamantane has NPA atomic charge of 0.955 on Ca, while 5F and 8F substitutions on CaO-adamantane lead to an increase of positive charge of 0.961 and 0.963, respectively.   

\begin{table}[ht]
\small
    \caption{Calculated FCFs, including Duschinsky rotations, bond-length changes from excited-ground state, and NPA charges of Ca and O atoms for various substituted CaO-adamantane ligands.}
    \label{tab:ca-ad}
  \begin{tabular}{l|llll}
    \hline
    Substituent  &  FCF & CaO-Change & Charge (Ca) & Charge (O)\\
    for H  & & (\AA) &  & \\
    \hline
    3NH$_{2}$ & 0.885 & -0.0226 & 0.9578 & -1.1335 \\
    (none) & 0.908 & -0.0234 &  0.9545 & -1.1458 \\
    3Cl & 0.929 & -0.0201 & 0.9610 & -1.1190 \\
    3F & 0.928 & -0.0203 & 0.9599 & -1.1230 \\
    5Cl & 0.939 & -0.0186 & 0.9622 & -1.1119 \\
    5F & 0.939 & -0.0188 & 0.9607 & -1.1157 \\
    8Cl & 0.914 & -0.0175 & 0.9655 & -1.1037 \\
    8F & 0.913 & -0.0168 & 0.9631 & -1.1066 \\
    \hline
  \end{tabular}
\end{table}

\begin{table}[ht]
\small
    \caption{Calculated FCFs, including Duschinsky rotations, bond-length changes from excited-ground state, and NPA charges for the Sr and O atoms, for various substituted SrO-adamantane ligands.}
    \label{tab:subad-FCF}
  \begin{tabular}{l | llll}
    \hline
    Substituent  &  FCF & SrO-Change & Charge (Sr) & Charge (O)\\
    for H  & & (\AA) &  & \\
    \hline
    3NH$_{2}$ & 0.872 & -0.0231 & 0.9694 & -1.1320 \\
    (none) & 0.873 & -0.0238 & 0.9710 & -1.1363 \\
    3Cl & 0.890 & -0.0211 &  0.9727 & -1.1177 \\
    3F & 0.898 & -0.0214 & 0.9715 & -1.1223 \\
    5Cl & 0.914 & -0.0203 & 0.9736 & -1.1106 \\
    5F & 0.915 & -0.0203 & 0.9722 & -1.1150\\
    8Cl & 0.889 & -0.0195 & 0.9770 & -1.1020 \\
    8F & 0.895 & -0.0187 & 0.9748 & -1.1054\\
    \hline
  \end{tabular}
\end{table}

While adding electron-withdrawing groups reliably increases the ionic character of the M-O bond, the corresponding increase in the FCF does not fully follow the trend, suggesting another factor besides bond ionicity being at play, as was also noticed by Ivanov et. al.\cite{Ivanov2020Toward}.  For example, the FCF increases up to 5F/5Cl subtitution, but decreases for 8F/8Cl substitution.  This is due to vibrational mode effects discussed next. 

\begin{figure}
\centering
  \includegraphics[width=0.45\textwidth]{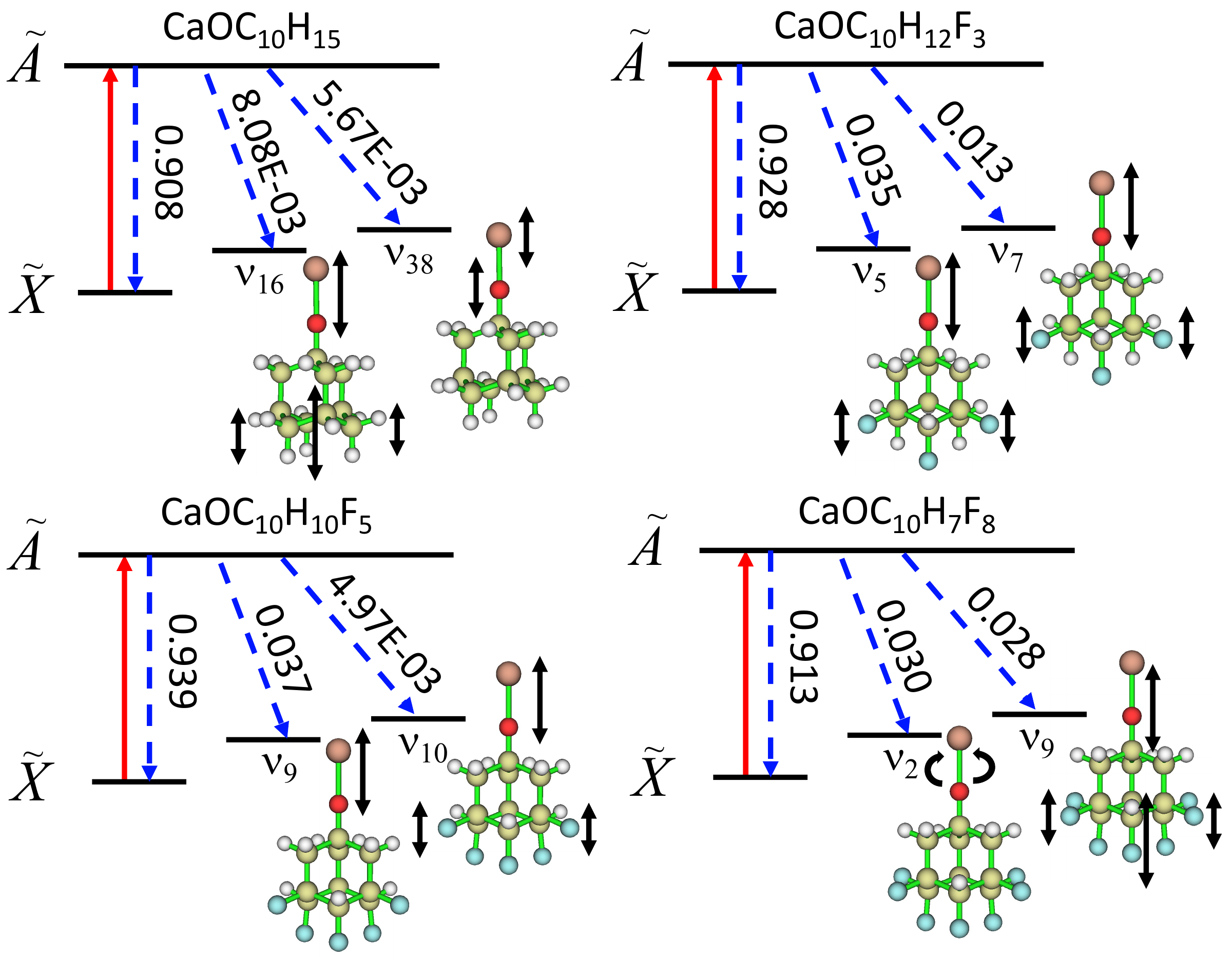}
  \caption{Excitation (red) and most dominant off-diagonal vibrational decays (blue) associated with each Franck-Condon factor for (top left to right): unsubstituted CaO-adamantane, 3F-CaO-adamantane, 5F-CaO-adamantane and 8F-CaO-adamantane.}
  \label{fig:Ca-ad}
\end{figure}

\begin{figure}
\centering
  \includegraphics[width=0.45\textwidth]{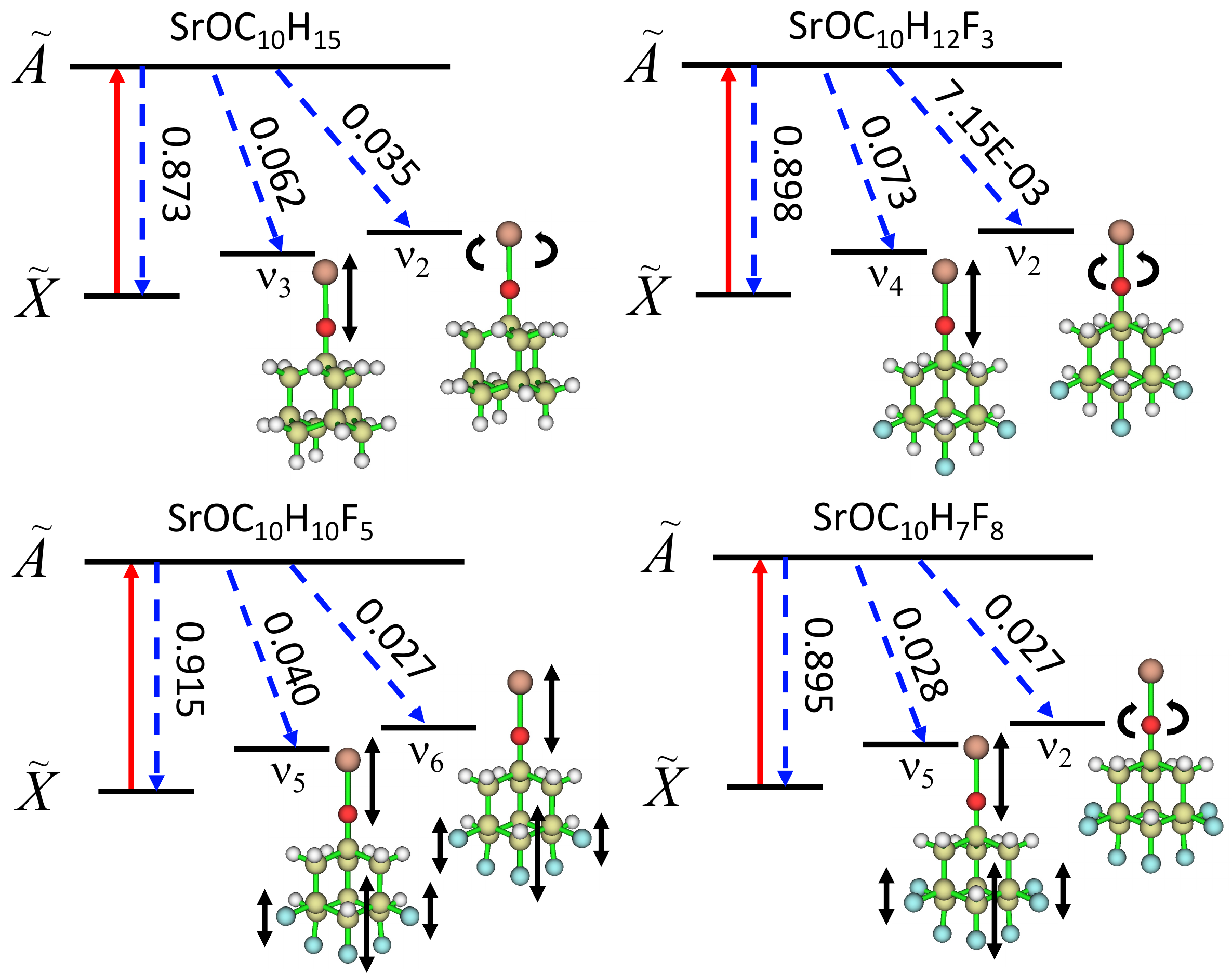}
  \caption{Excitation (red) and most dominant off-diagonal vibrational decays (blue) associated with each Franck-Condon factor for (top left to right): unsubstituted SrO-adamantane, 3F-SrO-adamantane, 5F-SrO-adamantane and 8F-SrO-adamantane.}
    \label{fig:Sr-ad}
\end{figure}

For both CaO-adamantane and SrO-adamantane, the behavior of the vibrational modes are similar (see Fig. \ref{fig:Ca-ad} and Fig. \ref{fig:Sr-ad}, respectively).  Figure \ref{fig:Ca-ad} shows the diagonal FCF and the largest two off-diagonal FCF leakage pathways for unsubstituted CaO-adamantane and all fluorine substitutions investigated (3F, 5F, 8F).  The off-diagonal FCF in CaO-adamantane is dominated by two vibrational modes, $\nu_{16}$ the Ca-O symmetric stretch, and $\nu_{38}$ the Ca-O-C asymmetric stretch.  As more electron-withdrawing substituents are added up to 5F/5Cl, the diagonal FCFs increase.  In 8F-substituted CaO-adamantane, the bending mode ($\nu_{2}$) dominates, and contributes on par with the symmetric stretch ($\nu_{9}$) that involves the motion of the ligand.  As more electron-withdrawing groups are placed on the molecule, the M-O stretch motion is more coupled with the ligand motion.  We hypothesize this coupling helps increase FCFs to a certain extent through symmetrizing the excited and ground state surfaces as in the case of 3F and 5F, but then too many withdrawing substituents encourage Ca-O bending motion, disrupting the dominant stretch PES and decreasing FCFs.

For SrO-adamantane, the picture is similar but slightly different (Fig. \ref{fig:Sr-ad}).  The bending mode, $\nu_{2}$, is always a dominanat off-diagonal decay, which decreases as substituents are added up to 5F, then reappears in 8F, which decreases FCFs.  We speculate this bending mode is more pronounced in Sr than Ca due to Sr's heavy mass changing the potential energy surface landscape.  Our calculations also lack spin-orbit coupling, which is also more pronounced in the Sr atom (though our theory was recently demonstrated to still predict Sr OCC trends in SrOPh in an upcoming paper).

Overall, these trends in FCFs can be used as a guide to predict optical cycling success; since they are less than $\approx 0.99$, they can be approximated as the vibrational branching ratio.

\section{Conclusions}
We demonstrate a direction for constructing larger R ligands, suitable for hosting quantum functional groups, Ca-O- and Sr-O-, for optical cycling.  Carbon-containing ligands (especially saturated) appear to be the most promising so far, due to their stability and large separation of bonding and antibonding states, away from the HOMO-LUMO OCC excitation of interest.  Moreover, adding electron-withdrawing substituents is shown to further increase FCFs via induction.  Lastly, symmetry and behavior of vibrational modes play an important role in controlling the diagonality of the OCC FCF.  Specifically, lower symmetry decreases FCFs, and substituents can change dominant mode behaviors in electronic decay pathways.  For our M-O-adamantane species, in which we design the M-O stretch to be the most dominant decay channel, electron-withdrawing substituents encourage ligand coupling to the M-O stretch mode, which symmetrizes the ground with the excited state potential energy surfaces, increasing FCFs, until electron-withdrawing effects overpower the PES symmetrizing benefits and encourages bending motions which disrupt the transition.  With these design rules in mind, larger molecules can be designed to have higher FCFs than smaller molecules, despite an increase in complexity and the number of vibrational modes. 

\emph{Acknowledgements --} This work was funded by the NSF Center for Chemical Innovation Phase I grant CHE-2221453.  CED acknowledges support from NSF grant DGE-2034835.  Computational resources at XSEDE and UCLA shared cluster hoffman2 are acknowledged.

\bibliography{arXiV}

\newpage 
~\newpage
\textbf{Supplementary Information}

\subsection{Benchmarking}

As seen in Table \ref{tab:tddft-ex}, the TD-DFT method correctly predicts vertical excitation energies, within 0.08 eV of the experimental band gaps for various large R group ligands.  Previously, CaOC$_{2}$H$_{5}$ was investigated theoretically via models from spectroscopic data\cite{Kozyryev2016Proposal} and with CASSCF.\cite{CASSCF-CaOC2H5}  

\begin{table}[ht]
\small
  \caption{Comparison between the computed vertical excitation energies (E$_{ex}^{vert}$, in eV), the computed adiabatic excitation energies (E$_{ex}^{ad}$, in eV) and available experimental band origins or excitation energies\cite{Brazier1986,Kozyryev2016Proposal,Guozhu2022CaOPh,Debayan2022} (T$_{e}$, in eV) for alkaline earth oxide derivatives excited from the X state to the A state.}
  \label{tab:tddft-ex}
  \begin{tabular}{l | lll}
    \hline
    Molecule & E$_{ex}^{vert}$ & E$_{ex}^{ad}$ & T$_{e}$ \\
    \hline
    CaOH & 1.973 & 1.969 & 1.979 \\
    CaOC$_{2}$H$_{5}$ & 1.956 & 1.951 & 1.965\\
    CaOPh & 1.924  & 1.922 & 2.010\\
    SrOPh & 1.779 & 1.776 & 1.854 \\
    CaO2Nap & 1.927 & 1.925 & 2.012 \\
    
    \hline
  \end{tabular}
\end{table}

\subsection{Potential Fitting}
\begin{figure}[ht]
    \centering
    \includegraphics[width=0.45\textwidth]{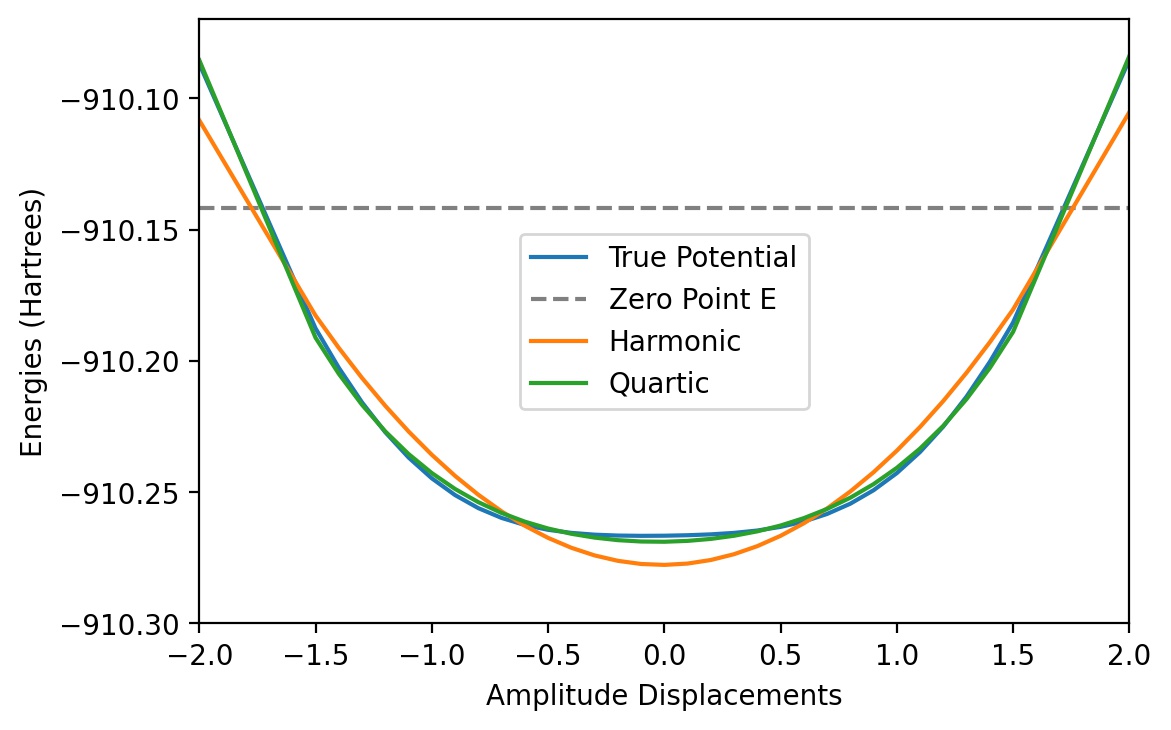}
    \caption{A harmonic vs. quartic potential fit to the potential energy surface scan along $\nu_{6}$ in the 1st excited state surface for CaOC$_{4}$H$_{9}$.}
    \label{fig:quartic}
\end{figure}

In CaOC$_{4}$H$_{9}$, the first two excited states are degenerate in energy and are likely strongly coupled.  We scanned the excited state potential energy surface along the largest off-diagonal decay mode, and fit a harmonic vs. quartic potential.  As seen in Fig. \ref{fig:quartic}, we find the quartic potential fits much better to this potential energy surface, suggesting a double-well minima (plot generated with MatPlotLib.\cite{Hunter:2007}).  

Since calculated FCFs use harmonic approximation and assume one minimum, we cannot predict correct FCFs for this molecule and two minima is a bad candidate for optical cycling.

Upon vertical excitation, the A and B electronic states are degenerate, but the ligand flexibility enables the Ca-R overlap and the splitting of the A and B states at the sides of the quartic well. The same effect does not hold true for CaO-adamantane because the ligand is rigid. 
\end{document}